\documentclass[final,1p,times]{elsarticle}
\biboptions{numbers,sort&compress}

\usepackage{amsmath}
\usepackage{xcolor}
\usepackage{amssymb}
\usepackage{float}
\usepackage{subcaption}
\usepackage{amsthm}
\usepackage{amssymb}
\usepackage{amsthm}
\usepackage{multirow}
\usepackage[bookmarks,bookmarksnumbered]{hyperref}
\hypersetup{colorlinks = true,linkcolor = blue,anchorcolor =red,citecolor = blue,filecolor = red,urlcolor = red,pdfauthor=author}
\usepackage{hyperref,graphicx}    
\begin{document}

\begin{frontmatter}


 \title{A New Approach in MRI Signal Processing for Detecting the Axonal Organization of the Brain}
 \author{Ashishi Puri\corref{cor1}\fnref{label1}}
 \ead{apuri@ma.iitr.ac.in}
 \author{Sanjeev Kumar\fnref{label1,label2}}
 \ead{sanjeev.kumar@ma.iitr.ac.in}
\cortext[cor1]{Corresponding author}
\address[label1]{Department of Mathematics,
Indian Institute of Technology Roorkee, Roorkee, 247667, India}
\address[label2]{Mehta Family School of Data Science and Artificial Intelligence, Indian Institute of Technology Roorkee, Roorkee, 247667, India}



\begin{abstract}
 This article introduces a new methodology for reconstructing the brain’s white matter fiber pathways in diffusion MRI. Usually, the signal intensity values will be lesser in the direction of higher diffusivity. The proposed approach picks the diffusion sensitivity gradient directions (dSGD), where the signal intensities are diminutive. Considering these as the directions of maximum diffusivity, we generate directions uniformly distributed around the picked dSGD. These newly computed uniformly spaced directions are considered gradient directions used in the reconstruction process. The state-of-art schemes like uniform gradient direction (UGD) have redundancy in the gradient direction, and adaptive gradient direction (AGD) has a constraint of solving linear system twice per voxel. These two limitations are turned down in this study simultaneously. Estimating gradient directions with the proposed scheme is employed in the multi-compartmental mixture models for calculating the fiber orientations. Simulation and experiments on the real data evaluate the feasibility of the proposed method.
\end{abstract}


\end{frontmatter}
\section{\textbf{Introduction}}
Diffusion tensor (DT) was initially put forward for use in magnetic resonance imaging (MRI) by Peter Basser in 1994 \cite{basser1994estimation, basser1994mr} called diffusion tensor imaging (DTI). It is an advanced non-invasive and in vivo technology that uses diffusion of water molecules for imaging the brain's white matter to provide the physiological and pathological condition of tissues of different body organs, especially the brain. An adequate literature can be found related to the fundamentals of DTI \cite{mori1999diffusion, luypaert2001diffusion, mori2006principles, jones2011diffusion}. \\

Magnetic field gradients are employed to create an image sensitized to diffusion in a specific direction for measuring the diffusion in MRI. A 3-dimensional diffusion model can be approximated to replicate this process of diffusion weighting for multiple varying diffusion sensitivity gradient directions (dSGD)  \cite{o2011introduction}. In a simplified way, DTI works by involving additional gradient pulses whose effect nullifies for immobilizing water molecules and produces a random phase shift for molecules that diffuse. The signals from these diffusing molecules are lost as a result of their random phase, which creates darker voxels (volumetric pixels) \cite{o2011introduction}. The white matter fiber (WMF) tracts parallel to the gradient direction appear dark in the diffusion-weighted image for that direction. The decreased signal $(S_k)$ is compared to the original signal $(S_0)$ for calculating the diffusion tensor $(D)$ by solving the Stejskal-Tanner (S-T) equation given in Eq. \ref{eq:Eq. 1}  \cite{basser1995inferring}. S-T equation shows how the signal intensity at each voxel declines in the presence of Gaussian diffusion:

\begin{equation} \label{eq:Eq. 1}
S_l=S_0e^{-b\hat{g}_l^TD\hat{g}_l}
\end{equation}

Here, $S_0$ is the image intensity measured without diffusion-sensitizing gradient and $S_k$ is the intensity measured  corresponding to $k^{th}$ diffusion-sensitizing gradient in unit direction $\hat{g}_k$ i.e. the dSGD. The product $\hat{g}_k^TD\hat{g}_k$ describes the diffusivity in direction $\hat{g}_k$. $b$ in Eq. \ref{eq:Eq. 1} is the LeBihan’s factor representing gradient strength, the pulse sequence and physical constants \cite{le1991molecular}. For a fixed value of $b$, higher diffusivity results in lower signal intensity and vice-versa. This means that dSGD corresponding to lower signal intensities shows the directions having maximum diffusion and hence loss of signal in that particular direction \cite{o2011introduction}.\\

Various models on a specific hypothesis are available for obtaining the linear system of equations for multi-fiber reconstruction. Some of the mixture models proposed till date are mixture of central Wishart distribution (MCW) \cite{jian2007multi, jian2007novel}, mixture of non-central Wishart distribution (MNCW) \cite{shakya2017multi}, mixture of von Mises-Fisher distributions \cite{kumar2008multi}, mixture of  Hyperspherical von Mises-Fisher \cite{kumar2009multi}, etc. Generally, these models function appropriately with large separation angles between two or three meeting/crossing fibers. However, the MNCW model gives the least angular error compared to the other models \cite{shakya2017multi}. In MNCW model, the non-negative least square method (NNLS) \cite{lawson1995solving} and advanced orthogonal matching pursuit (OMP) algorithm called OMP-TV2 are used to solve the linear system in \cite{puri2021enhanced} and \cite{puri2022omp}, respectively.\\

In \cite{puri2021enhanced}, uniform gradient directions (UGD) and adaptive gradient directions (AGD) approaches are discussed. In the UGD approach, a fixed and larger number of gradient directions (uniformly distributed on unit sphere or hemisphere) are used to reconstruct white matter fibers. In the AGD approach, we initially use very few gradient directions to obtain a rough idea about the fiber orientations. Subsequently, we use these rough fiber orientations to generate new gradient directions so that new gradient directions are uniformly distributed around the rough fiber direction. Among multiple gradient directions that are used in reconstruction, directions representing original fiber orientations are of prime importance. Choosing the larger number of gradient directions  (like in the UGD approach) results in an enormous number of unnecessary directions for a particular voxel. AGD approach \cite{puri2021enhanced} is introduced because of these artifacts. AGD approach successfully eliminates this issue and also reduces the angular error. Instead of taking a large number of gradient directions, a comparatively small number is chosen initially followed by using a new set of AGD. In particular, emphasis is precisely given to the directions where fibers are highly anticipated. In the AGD approach, the linear system of equations is required to be solved twice in each voxel to procure the outcome of the fiber orientations. First, we solve the linear system for a smaller number of gradient directions, followed by again for the updated gradient directions.  \\

In this study, we have worked on all these limitations. The proposed approach eliminates the issue of using redundant directions and also solving linear systems twice per voxel. Here, we scan the dSGD array and pick the directions corresponding to which maximum diffusion is observed. Then, we generate the gradient directions that are uniformly distributed around the dSGD picked earlier. Moreover, the dSGD is the centre of the generated gradient directions. These generated directions then work as the gradient directions for the reconstruction process. Therefore,  we must solve the linear system only once per voxel and reduce the redundancy in the gradient directions. Multiple simulations and experiments on real datasets are performed to support the proposed idea's authenticity. The simulated experiments are performed at diverse noise levels to evaluate the robustness of the model. The proposed model performed well enough in reducing the angular errors while detecting the brain's axonal (white matter) organization.

\section{Methods}
\subsection{Linear system of equations}

The signal equation based on non-central Wishart distribution \cite{james1955non, letac2004tutorial, li2003noncentral} is given as follows \cite{shakya2017multi}:

\begin{equation} \label{eq:Eq. 2}
S_l/S_0= \sum_{m=1}^K w_m(1+trace(\Sigma_m\textbf{B}_l))^{-q}\exp~[-trace(\textbf{B}_l\{I_p+\Sigma_m\textbf{B}_l\}^{-1}\Omega_m)]
\end{equation}

where $\textbf{B}=b\hat{g}_l\hat{g}_l^T$ and $\hat{g}_l$ is the dSGD used for generating the signals such that $S_l$ is the signal intensity along $\hat{g}_l^{th}$ direction. Using the expected value of non-central Wishart distribution $(\textbf{D} = q\Sigma+\Omega)$, we have $\textbf{D}_m=q\Sigma_m + \Omega_m$ \cite{shakya2017multi}. Reader is referred to \cite{shakya2017multi} for the further information about all the parameters involved in this model. \\

Eq. \ref{eq:Eq. 2} gives rise to following linear system of equations:

\begin{equation} \label{eq:Eq. 3}
A\textbf{w}=y+\eta
\end{equation}

where $y=S_l/S_0$ is the normalized signal vector, $\eta$ is the noise. Each entry of matrix $A$ is given as,\\
 
\hspace{3cm}$A_{lm}= (1   + trace(\Sigma_{m}{\textbf{B}}_l))^{-q}\exp~[-trace\{\textbf{B}_l(I_p + \Sigma_m \textbf{B}_l)^{-1}
\Omega_m\}$]\\

where, $l=1,2,...,N$ are DTI measurements. The vector $\textbf{w} = \{(w_m): m=1,2,...,K\}$ in Eq. \ref{eq:Eq. 3}, is the unknown parameter that has to be calculated. Here, $K$ denotes the mixture components/compartments of each voxel or the number of gradient directions used in reconstruction process and $w_m$ is the mixture weights corresponding to the $m^{th}$ gradient directions.\\

\subsection{Adaptive gradient directions approach}

\begin{figure}[H]
  \centering
 {\includegraphics[width=7in]{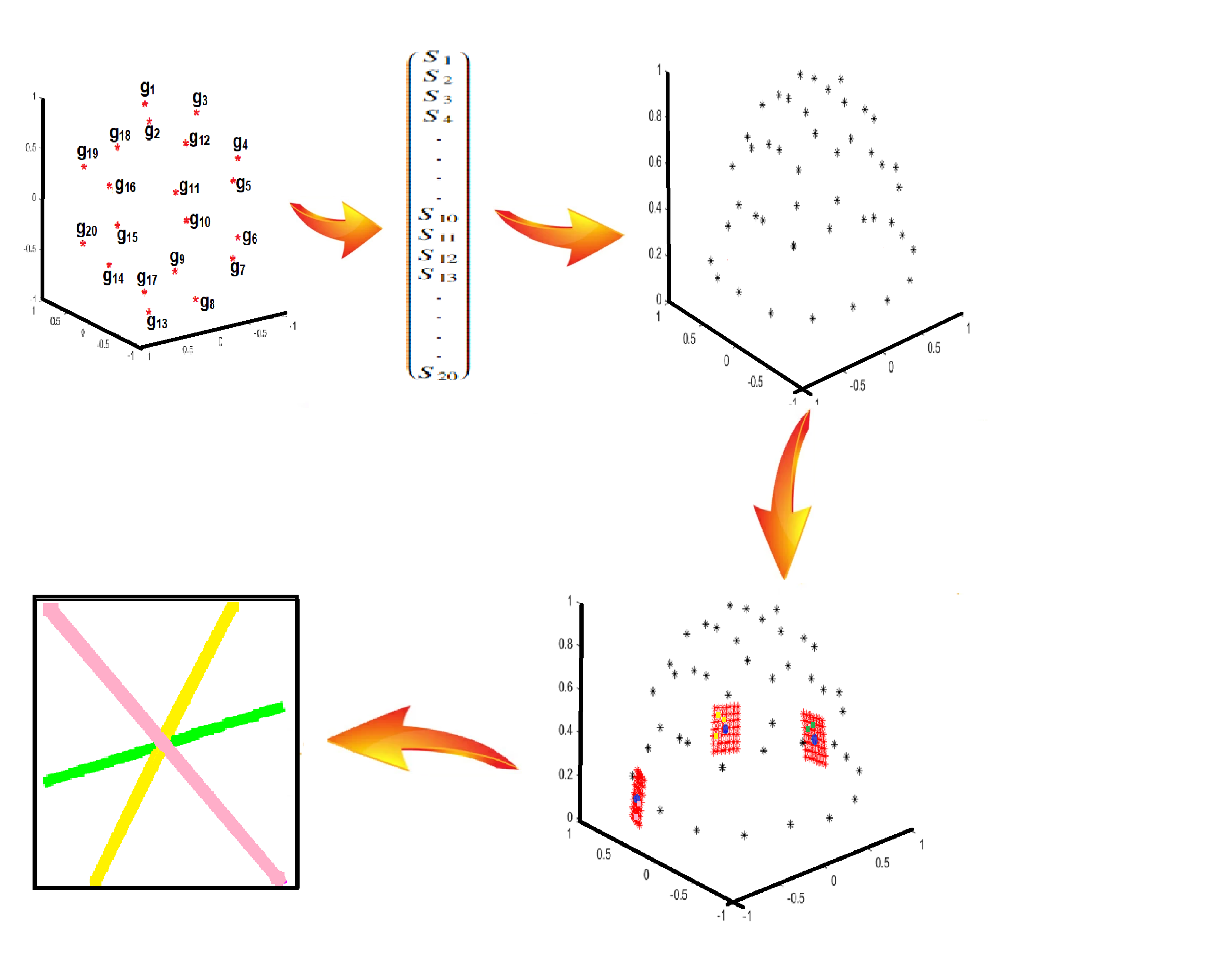}}
  \caption{Schematic diagram representing the working process of the AGD model.}
  \label{fig:1}
\end{figure}
In this approach, we do not play with the dSGD. However, gradient directions used in the reconstruction process are taken into consideration in this approach. The dSGD generates signals and these signals are then used in Eq. \ref{eq:Eq. 2} as $S_l's$. The `$l$' dSGD used for generating signals are shown in the first image of  Fig. \ref{fig:1} with red stars, and `$K$' uniform gradient directions over a unit hemisphere (for reconstructing) indicated by black stars are shown in the third image. All the red stars representing dSGD ($l=20$) are indicated by $g_1,~g_2,~g_3,~.~.~.~,g_{20}$. Using these two sets of directions, we solve Eq. \ref{eq:Eq. 3} and obtain the weight vector `$\textbf{w}$' for the first step of the AGD approach. `$A$' becomes a matrix of size $N\times K$. NNLS method is employed to obtain a sparse weight vector of size $K\times 1$. The positions of non-zero values are noted, and gradient directions (among `$K$' directions) at these positions indicate the rough fiber orientations in first step. In the subsequent step, updated gradient directions for reconstruction are used in place of $K$ gradient directions. These updated gradient directions are shown as red stars in form of grids in fourth image of Fig. \ref{fig:1}. Let $K'$ be the number of new gradient directions for second step, hence size of matrix `$A$' becomes $N\times K'$ . Again, NNLS method is employed to obtain sparse weight vector such that gradient directions corresponding to non-zero positions in `$\textbf{w}$' are the direction of expected fibers.\\

 Now, we calculate the average of azimuthal and polar angles of fibers which are very closely oriented such that they amalgamate to become one fiber. The average of azimuthal and polar angles of two vectors represented with pink dots inside the grid is taken to obtain the final reconstructed pink fiber. Similarly, three yellow dots are considered very closely oriented, hence average of their azimuthal and polar angles is taken to obtain final yellow colored fiber. Following the same way, green colored fiber is reconstructed.

\subsection{Proposed work}
\begin{figure}[H]
  \centering
  {\includegraphics[width=6.4in]{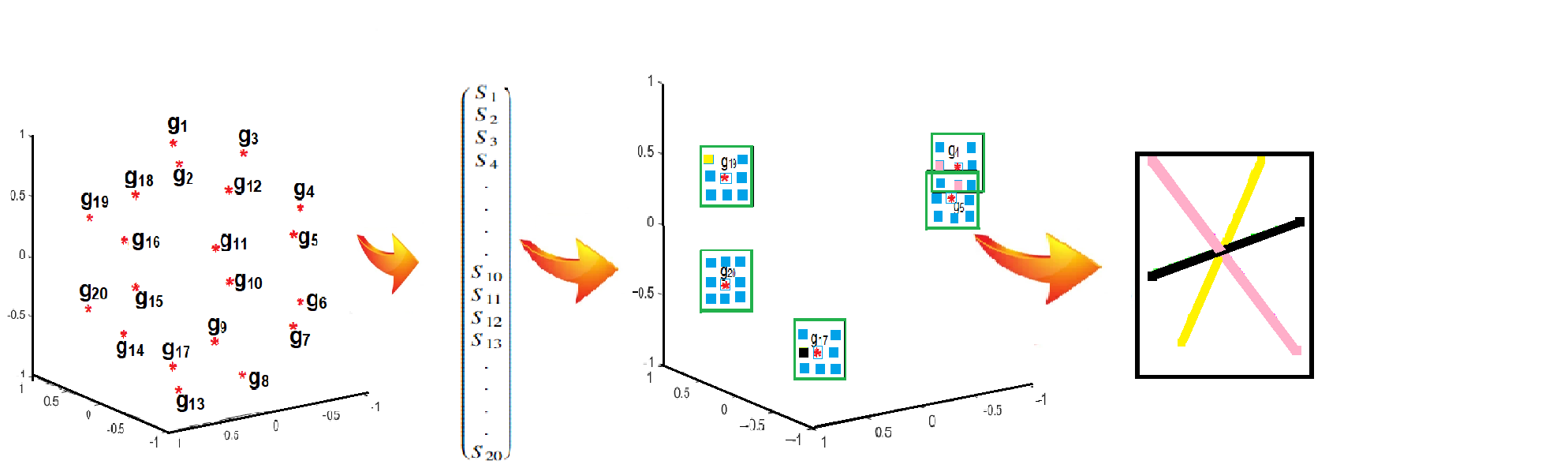}}
  \caption{Schematic diagram representing the working process of the proposed model.}
  \label{fig:2}
\end{figure}

In this study, we generate signals corresponding to the dSGD (uniformly distributed over a unit sphere or hemisphere) in the simulations. Moreover, the dSGD and the corresponding signal intensities are available to us in the case of a real dataset. We begin our algorithm by analysing the signal intensity vector and picking a certain number of signal intensities with the least values. Noting the indices of these values, we select the corresponding dSGD. These selected directions are shown with red stars inside the grids in the third image in Fig. \ref{fig:2}. We precisely work along the directions nearer to red stars inside the grids. New gradient directions are generated uniformly around the selected dSGD such that multiple grids \cite{puri2021enhanced} are formed on the surface of a unit hemisphere. The blue markers represent the generated gradient directions inside the grids in Fig. \ref{fig:2}. All these directions together work as the `$K$' gradient directions in Eq. \ref{eq:Eq. 3}. In AGD approach, the linear system obtained in Eq. \ref{eq:Eq. 3} needs to be solved twice for each voxel. In the proposed approach, we mitigate the strive of solving the linear system twice. Here, the system is solved once for each voxel considering the blue markers as the `$K$' directions for Eq. \ref{eq:Eq. 3}.

\subsection{Step-wise Description of the Proposed Algorithm}

\textbf{Step-1:} Picking `$k$' signal intensities with least values followed by noting the indices of these values in the signal vector, we select the corresponding dSGD vectors. These selected vectors are shown as red stars inside the grids in Fig. \ref{fig:2}.\\

\textbf{Step-2:} Next, convert these vectors from Cartesian to the spherical coordinate system. Therefore, we obtain `$k$' (say) pairs of polar and azimuthal angle. \\

\textbf{Step-3:} For each pair, we have spherical coordinates $(\phi_{n},\theta_{n},r)$ with fixed radial distance, $r=1$ and  $k \in \mathbb{Z}^{+}$, where $n=1,2,3,...,k$. \\

\textbf{Step-4:} Now, we generate uniformly distributed gradient directions in the neighbourhood of `$k$' points obtained in the above step. These points are on the hemisphere's surface, forming a grid centred at each $n^{th}$ orientation. New vectors/directions are computed in a spherical coordinate system as follows:\\

\hspace{0.5cm} \textbf{Step-4.1:} Collection of new computed azimuthal angles = $\{\phi_{n},~\phi_{n} \pm m_1, ~\phi_{n} \pm 2m_1, .~.~.~, \phi_{n} \pm c_1m_1\}$
such $c_1, m_1 \in \mathbb{Z}^+$ where $c_1$ accounts for the number of points inside the grid and $m_1$ accounts for the spacing between the points.\\

\hspace{0.5cm} \textbf{Step-4.2:}  Collection of new computed polar angles = $\{\theta_{n},~\theta_{n} \pm m_1, ~\theta_{n} \pm 2m_1, .~.~.~, \theta_{n} \pm c_1m_1\}$.\\
 
\hspace{0.5cm} \textbf{Step-4.3:} Collection of computed pairs of azimuthal and polar angles   = $\bigg\{(\phi_{n},\theta_{n}), (\phi_{n}, \theta_{n} \pm m_1),~.~.~.~,~$

$(\phi_{n},\theta_{n}\pm c_1m_1),~(\phi_{n}\pm m_1,\theta_{n}),~(\phi_{n}\pm m_1,\theta_{n} \pm m_1),~ .~ .~ .~ ,~(\phi_{n}\pm m_1,\theta_{n} \pm c_1m_1),~ .~ .~ .~ ,~$

$(\phi_{n}\pm c_1m_1,\theta_{n}),~(\phi_{n}\pm c_1m_1,\theta_{n} \pm m_1), ~. ~. ~.~, ~ ~(\phi_{n}\pm c_1m_1,\theta_{n} \pm c_1m_1)\bigg\} $\\

The total number of new directions computed per grid = $(2c_1+1)^2$\\
The total number of GD obtained i.e.  $K$ = $(2c_1+1)^2\times k$\\

\textbf{Step-5:} Next, converting these $K$ spherical coordinates with radial distance unity to a Cartesian coordinate system and these vectors will then replace the $K$ vector required in Eq. \ref{eq:Eq. 3}. These points are represented by the blue markers inside grids in Fig. \ref{fig:2}. 

\section{Results and Discussion}
In simulations, $K=46$ diffusion sensitizing gradient directions are considered for generating the signal vector ($S$) of size $46\times1$. Following \cite{jian2007multi,shakya2017multi,puri2021enhanced}, we set $q=2$ ($q$ $\in$ Gindikin ensemble $\Lambda=\{\frac{a}{2}:a=1,2,...,(p-2)\}\cup\big[\frac{p-1}{2},\infty))$ \cite{gindikin1975invariant, n1988davidson, peddada1991proof}, $b=1500s/mm^2$. We employ non-negative least square
method (NNLS) \cite{lawson1995solving} and OMP-TV2 algorithm \cite{puri2022omp} to solve the linear system of equations given in Eq. \ref{eq:Eq. 3} in the proposed work for calculating the weight vector, $\textbf{w}$. Synthetic data is generated using MATLAB$^{TM}$ open library \cite{barmpoutis2009adaptive, barmpoutis2010tutorial}. Resultant angular error (RAE) is considered as evaluation metric for comparing different models. RAE is defined as the sum of angular errors corresponding to azimuthal and polar angle. In simulations, polar angle is assigned a fixed value of $90^\circ$. RAE is calculate as follows:

$$A.E. = \frac{\sum_{k=1}^{I}[\lvert \phi_k^{original}- \phi_k^{estimated}\rvert + \lvert \theta_k^{original}- \theta_k^{estimated}\rvert]}{I}$$

where, $I$ denotes total number of fibers per voxel. Rician distributed (RD) noise is added to signal vector $(y)$ considering the standard deviation of noise denoted by $\sigma$, signal vector become:

$$y=\sqrt{(y+\sigma*randn(1))^2+(\sigma*randn(1))^2},$$
\vspace{0.1cm}

where, function `randn' generates random numbers that are distributed normally. In this article, simulations have been primarily performed with RD noise, $\sigma=0.01-0.1$.\\

\subsection{Simulation Results}

\begin{figure}[ht]
  \centering
  \subcaptionbox{\label{fig3:a}}{\includegraphics[width=2.3in]{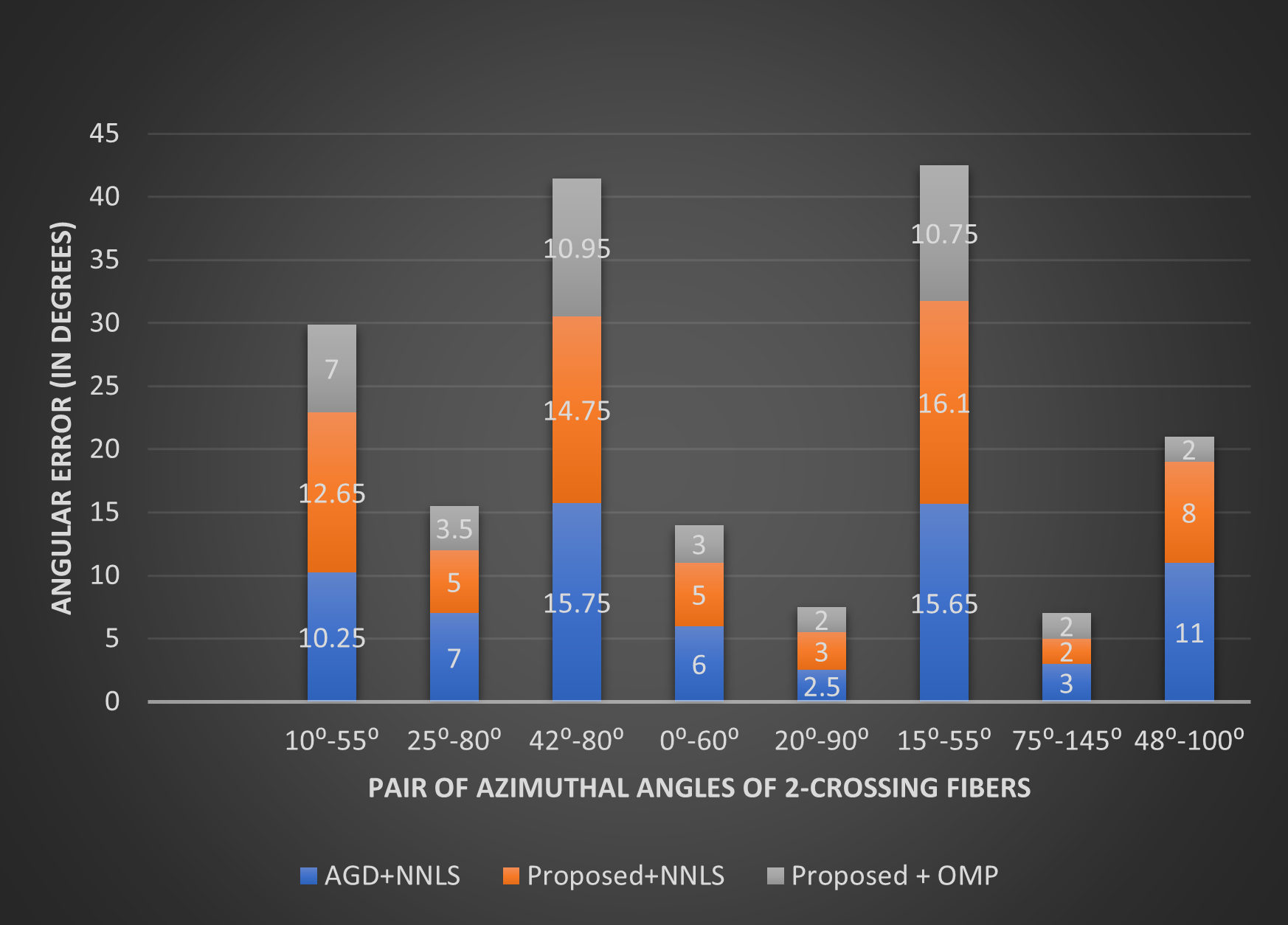}}\hfill
  \subcaptionbox{\label{fig3:b}}{\includegraphics[width=2.3in]{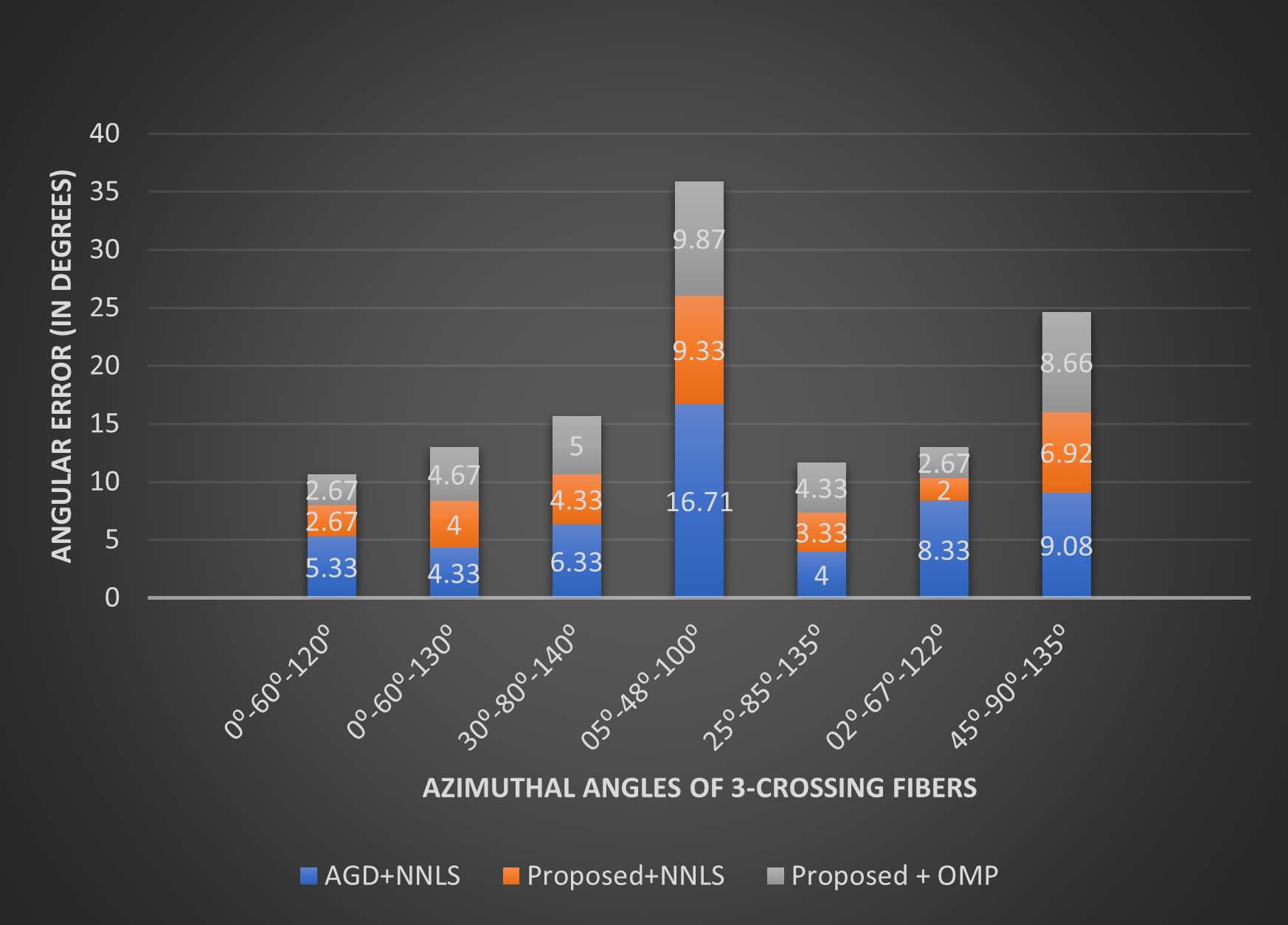}}
  \caption{Resultant angular error using proposed model and AGD model in case of (a) 2-crossing fibers and (b) 3-crossing fibers per voxel at $\sigma=0.05$.}
  \label{fig:3}
\end{figure}

We have carried out a detailed simulation study using the proposed approach. The plots shown in Fig. \ref{fig3:a} and Fig. \ref{fig3:b}  represent the performance of the AGD model (with NNLS) and the proposed model (with both NNLS and OMP-TV2). All these experiments are done with $\sigma=0.05$. Fig. \ref{fig3:a} illustrates the RAE for 2-crossing fibers aligned in varying angles, whereas, the Fig. \ref{fig3:b} represents a plot of RAE in case of 3-crossing fibers. Since, the proposed approach performs better when coupled with OMP-TV2, hence, visual results are shown with OMP-TV2 coupled with proposed approach in the further experiments.

\begin{figure}[ht]
  \centering
  \subcaptionbox{\label{fig4:a}}{\includegraphics[width=2.3in]{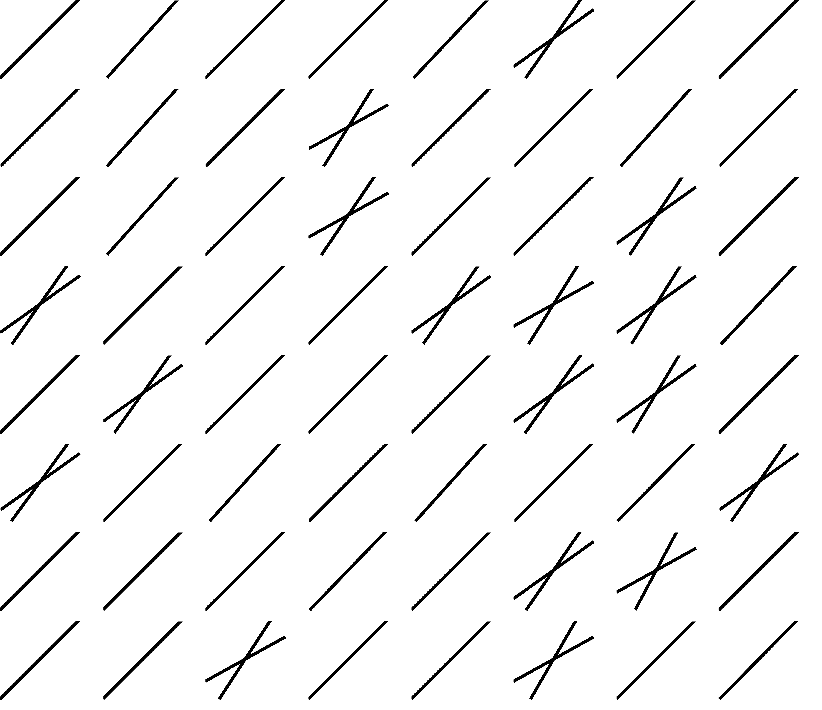}}\hfill
  \subcaptionbox{\label{fig4:b}}{\includegraphics[width=2.3in]{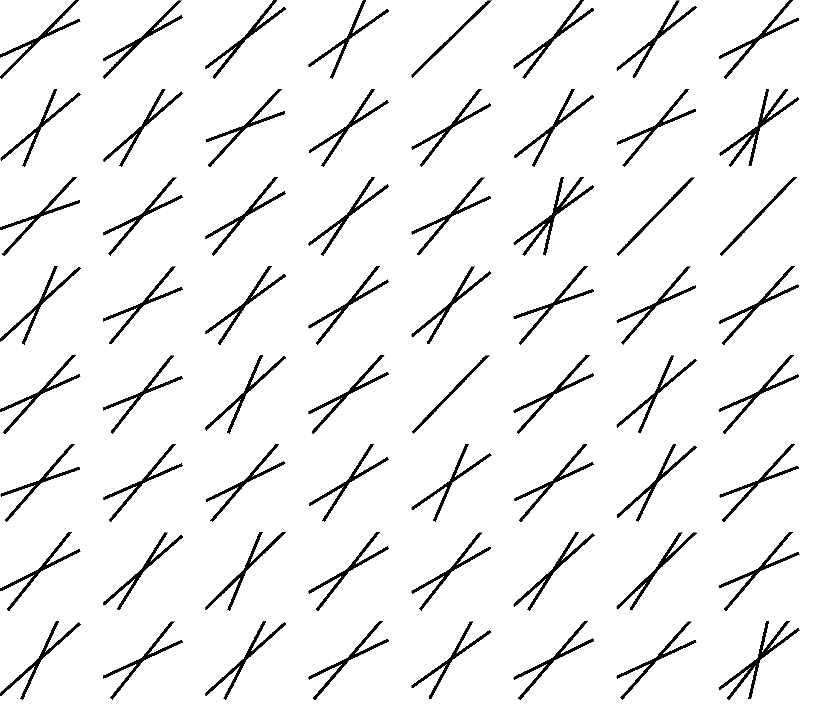}}
  \caption{Reconstruction of 2-crossing fibers with separation angle of $40^\circ$ and noise, $\sigma=0.05$ using (a) AGD  model and (b) proposed model, respectively.}
  \label{fig:4}
\end{figure}

\begin{figure}[ht]
  \centering
  \subcaptionbox{\label{fig5:a}}{\includegraphics[width=2.3in]{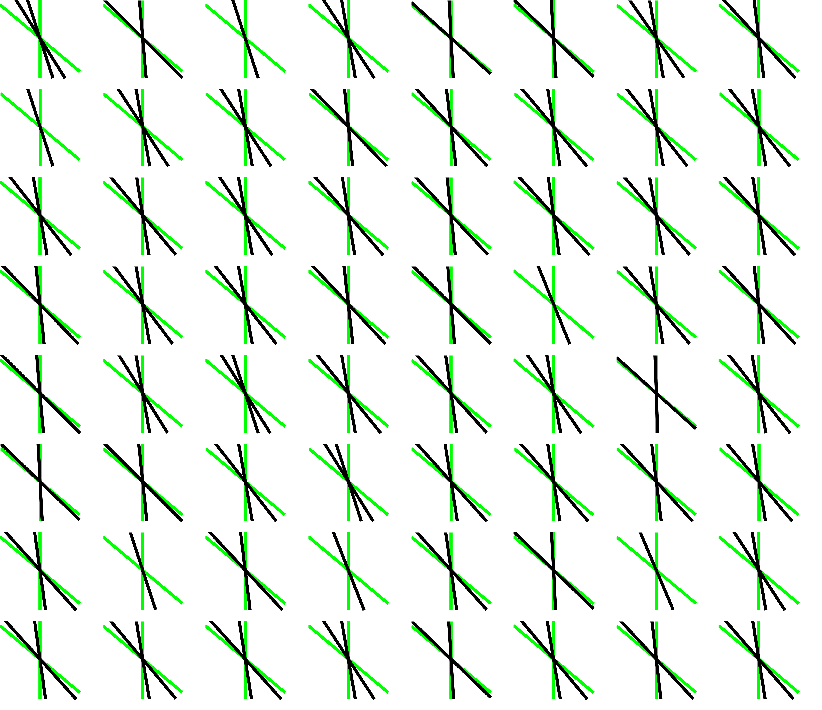}}\hfill
  \subcaptionbox{\label{fig5:b}}{\includegraphics[width=2.3in]{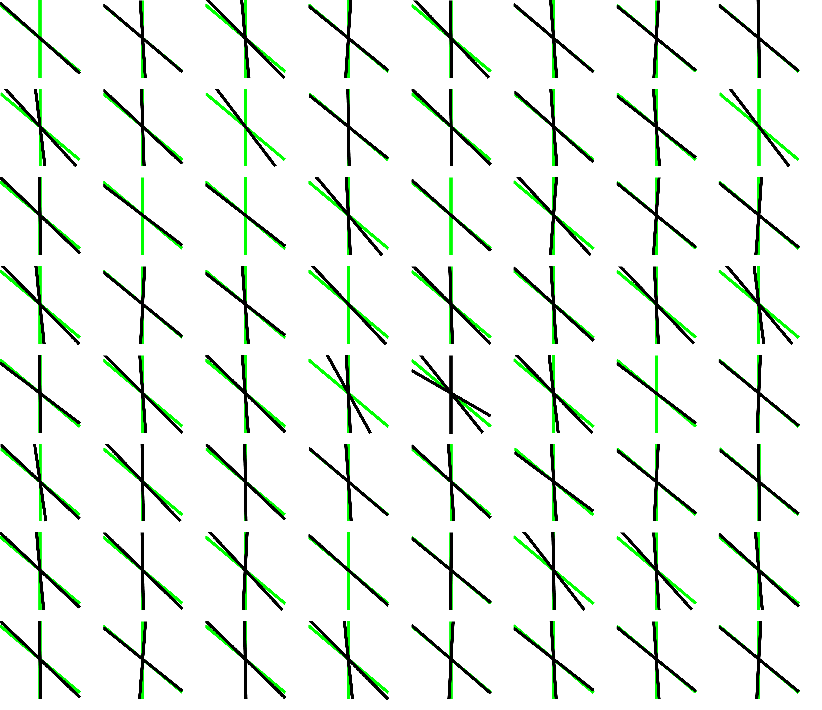}}

  \caption{Reconstruction of 2-crossing fibers with separation angle of $50^\circ$ and noise, $\sigma=0.06$  using (a) AGD  model and (b) proposed model, respectively.}
  \label{fig:5}
\end{figure}

 We performed simulations with a pair of fibers per voxel. These fibers are oriented such that the separation angle between them is small i.e. $\lvert \phi_1-\phi_2\rvert=40^\circ$ and dataset is equipped with noise, $\sigma=0.05$. A total of $64$ simulations are performed and shown in Fig \ref{fig:4}. The reconstructed results (images) using AGD approach and the proposed approach are shown in Fig. \ref{fig4:a} and Fig. \ref{fig4:b}, respectively. It is fairly easy to perceive and visualize which model works better as the AGD model is not reconstructing two crossing fibers appropriately. Infact, in almost all the voxels, a single fiber is detected, whereas the proposed model is capable of performing well and detecting two fibers in almost $90\%$ of the voxels. Secondly, we simulated an experiment for 2-crossing fibers per voxel aligned such that $(\phi_1,\phi_2)=(90^\circ,140^\circ)$ and noise, $\sigma=0.06$. Although both the models detect the presence of two fibers crossing each other, it is important to see which model detect orientation of the fibers correctly. Hence, to visualize and compare the two models' accuracy, we overlapped the original fiber directions with the reconstructed ones in each voxel as shown in Fig. \ref{fig:5}. This makes it easier to comprehend the accuracy of the proposed model. The original fibers are shown using green colour, and the reconstructed fibers with black colour. We can observe that in case of AGD model the reconstructed fibers are not completely overlapping the ground-truth fibers but in case of proposed model, except 2-3 voxels, the green and black fibers are overlapping. Hence, this represents that among proposed model and AGD model, former performs better in reconstructing the fibers with higher accuracy. 
 
\begin{figure}[ht]
  \centering
  \subcaptionbox{\label{fig6:a}}{\includegraphics[width=2.3in]{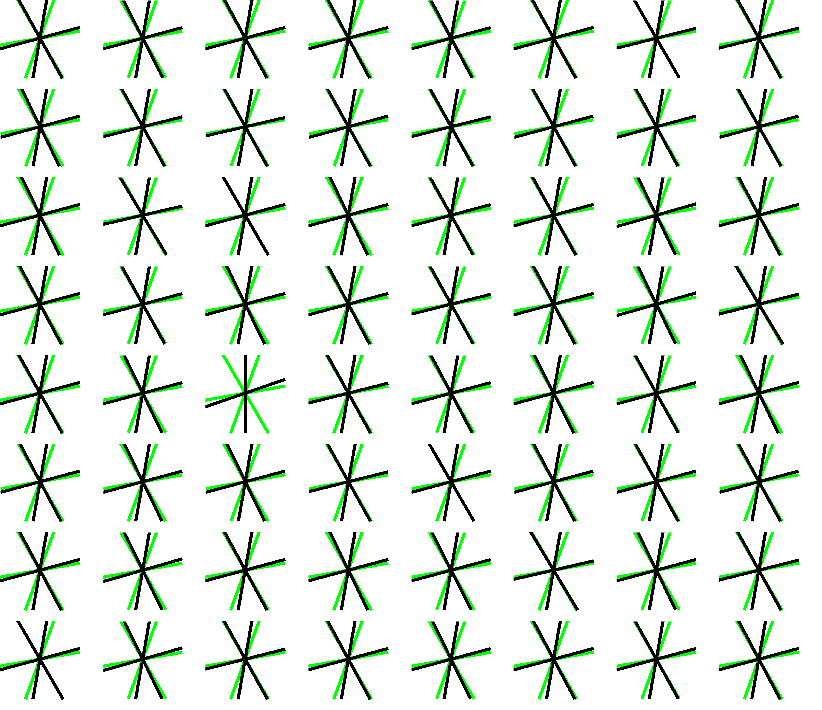}}\hfill
  \subcaptionbox{\label{fig6:b}}{\includegraphics[width=2.3in]{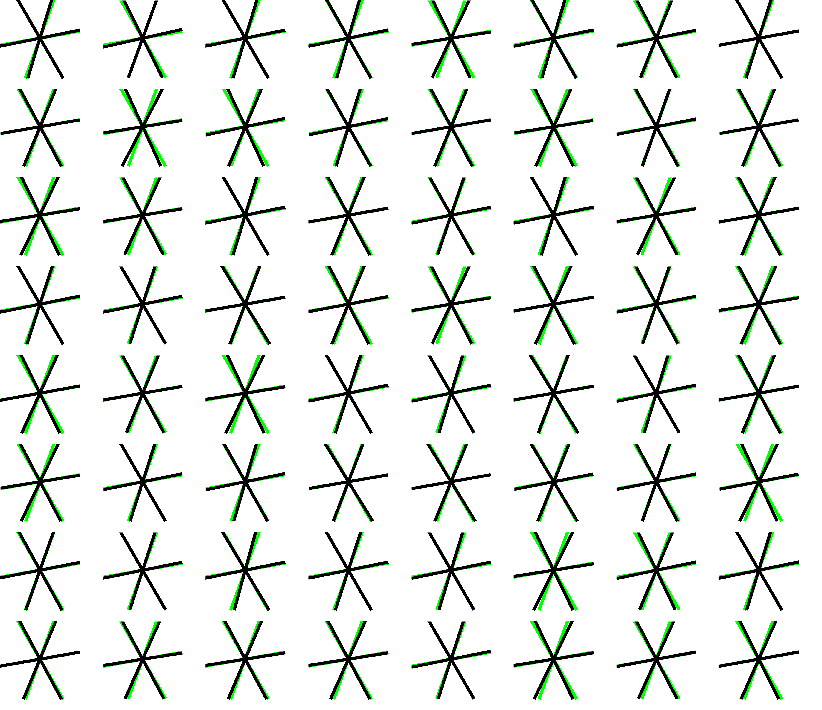}}

  \caption{Reconstruction of 3-crossing fibers with original orientation as $(\phi_2,\phi_2,\phi_3)= (10^\circ,70^\circ,120^\circ)$ and noise, $\sigma=0.1$  using (a) AGD  model and (b) proposed model, respectively. }
  \label{fig:6}
\end{figure}

Proceeding to the 3-crossing fibers simulations, we took $(\phi_1,\phi_2,\phi_3)=(10^\circ,70^\circ,120^\circ)$ and a higher level of noise, $\sigma=0.1$, is introduced in the dataset. Again, the ground-truth fiber directions are overlapped with the reconstructed ones as mentioned above. It assists in understanding the performance of both models with more clarity. Although both models perform well in detecting the accurate number of fibers i.e. 3-crossing fibers in each voxel, the proposed model's accuracy is better. As shown in Fig. \ref{fig:6}, the reconstructed fibers are completely overlapping the ground-truth fibers in all the voxel using proposed model but AGD image shows slight disorientation of fibers in nearly all the voxel. This dataset was highly noisy, hence the proposed model is showing its robustness by performing good at different noise levels.

\begin{figure}[ht]
  \centering
  \subcaptionbox{\label{fig7:a}}{\includegraphics[width=2.3in]{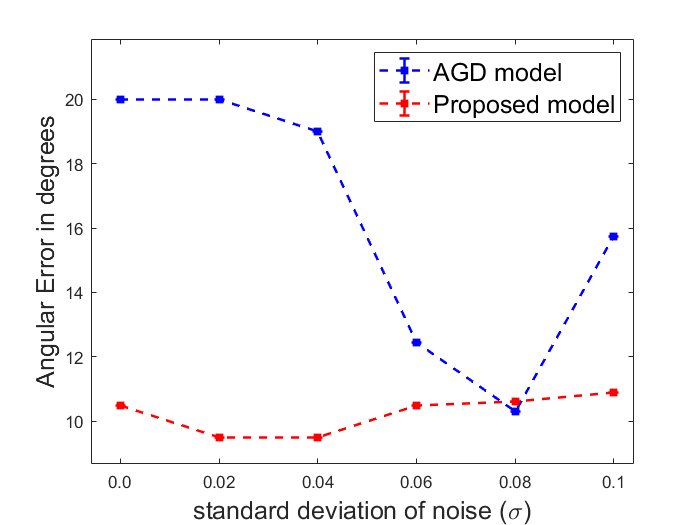}}\hfill
  \subcaptionbox{\label{fig7:b}}{\includegraphics[width=2.3in]{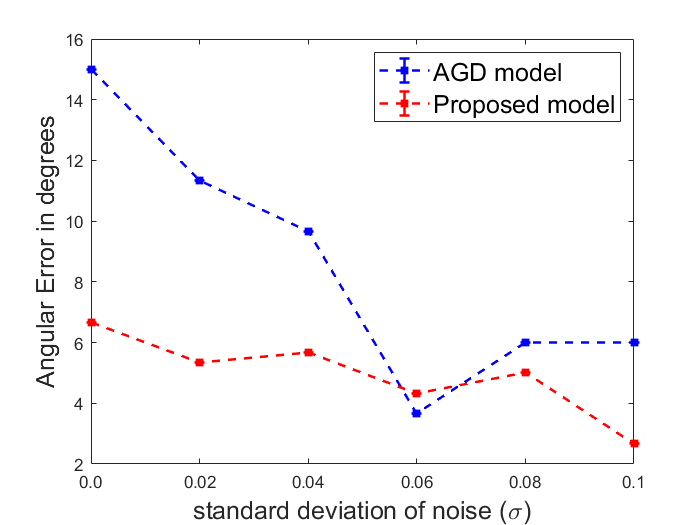}}

  \caption{Plot representing mean angular error for (a) 2-crossing fibers with $(\phi_1,\phi_2)=(25^\circ,65^\circ)$ and (b) 3-crossing fibers with $(\phi_1,\phi_2,\phi_3)=(5^\circ,65^\circ,115^\circ)$ at varying noise noise levels. }
  \label{fig:7}
\end{figure}

In simulations, we analyze the performances of both approaches using noisy signals. We examine a case of 2-fibers oriented such that $\phi_1=25^\circ$ and $\phi_2=65^\circ$ in Fig. \ref{fig7:a}. A small separation angle between the fibers is a good choice for analyzing the significance of the proposed model. This is because the state-of-art models mainly has constraints and limitations at smaller separation angles. In parallel, we took 3-fibers with orientations $(\phi_1,\phi_2,\phi_3)=(5^\circ,65^\circ,115^\circ)$  in Fig. \ref{fig7:b}. The RAE using both the models is shown  in Fig. \ref{fig:7}. In both cases, we see that the proposed model performs better at every noise level.

\subsection{Experiments on real data}
\begin{figure}[H]
  \centering
  {\includegraphics[width=3in]{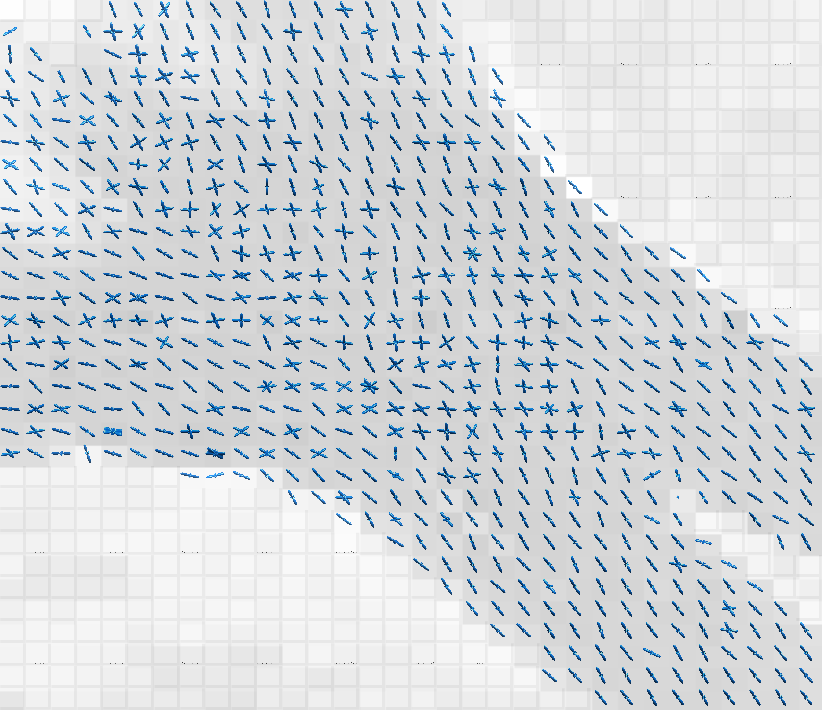}}
  \caption{Reconstruction of orientational map of optic chiasm of rat's brain using proposed model.}
  \label{fig:8}
\end{figure}

We applied the proposed model to two different
sets of real data. We started with a rat optic chiasm DW-MRI data set which is a single-shell dataset. The optic chiasm, or optic chiasma, is that fragment of the brain where the optic nerves intersect. Hence, it is of primary importance to the visual tracks. It is located at the base of the brain inferior to the hypothalamus and approximately $10 mm$ superior to the pituitary gland within the suprasellar cistern \cite{ireland2022neuroanatomy}. This dataset can be accessed freely at \cite{barmpoutis2010tutorial}. The images and signals were acquired for 46 directions with a $b$-value of $1250 s/mm^2$. A single image was also acquired at $b \approx 0 s/mm^2$. Other parameters of the experiment include Echo time (TE), Repetition time (TR), $\Delta$ and $\delta$ taken as $25 ms, 1.17 s, 17.5 ms, 1.5 ms$, respectively. The reconstructed image obtained using the proposed model is overlapped with the reference image are shown in Fig. \ref{fig:8}. Following \cite{puri2021enhanced,puri2022omp,barmpoutis2009adaptive,kumar2008multi,kumar2009multi,shakya2017multi} and the definition of optic chiasm, it is anticipated that the reconstructed image must indicate two fiber bundles coming in some pattern to intersect in the middle. Our model can detect
the orientations of optic nerves and their crossings at the centre region, although the
directions of crossing fibers are not always coherent.\\

We also carried an experiment on a four-shell dataset \cite{fan2014investigating, fan2016mgh}. The dataset was
collected with the customized Siemens 3T Connectom scanner (MAGNETOM Skyra
Siemens Healthcare) for healthy adults and downloaded from Human Connectome Project. Diffusion scans were taken at four
different $b$-values of $1000, 3000, 5000$ and $10000 s/mm^2$
and the number of diffusion sensitizing gradient directions were 64, 64, 128 and two sets of 128 for the four $b$-values, respectively. The details for this dataset can be seen at \cite{fan2014investigating, fan2016mgh}. We have chosen $50^{th}$ slice in our experiments. Here, the value of $N$ is 552 for the dataset. The data matrix size is $128 \times 128$ with 78 slices, the resolution is $1.74\times 1.74\times1.7 mm^3$ and the values of TR, TE and multiband factor are  5000 ms, 108 ms and 3, respectively.\\ 

\begin{figure}[!ht]
  \centering
  \subcaptionbox{\label{fig9:a}}{\includegraphics[width=2in]{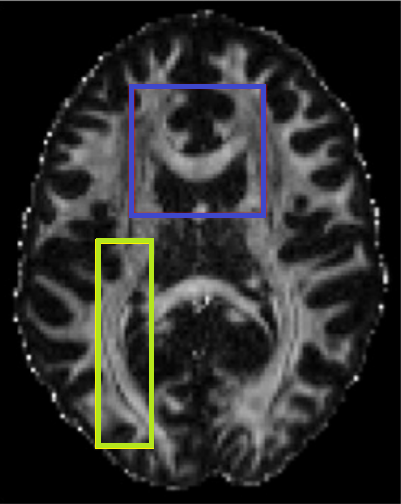}}\hfill
   \subcaptionbox{\label{fig9:b}}{\includegraphics[width=2in]{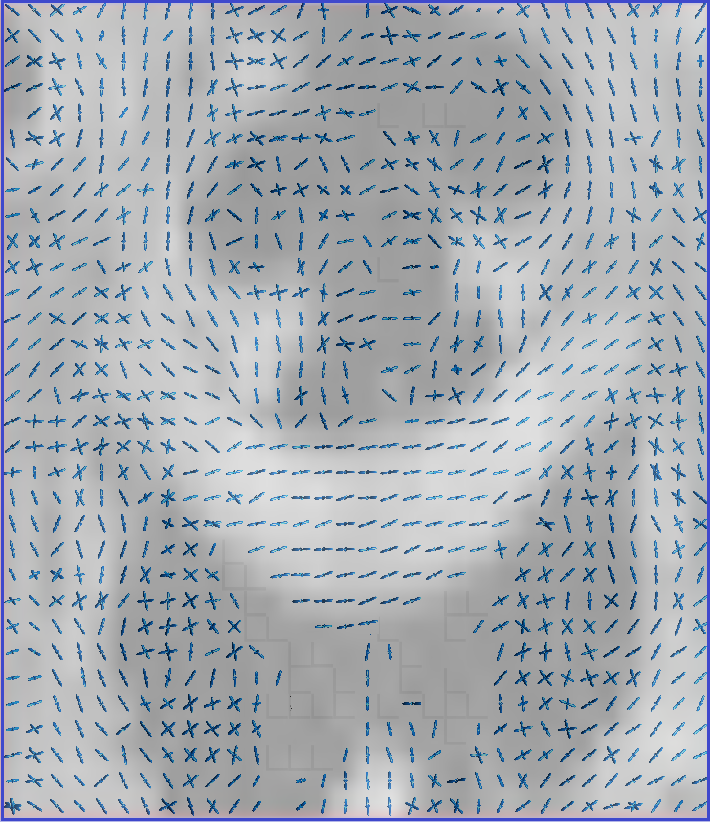}}\hfill
  \subcaptionbox{\label{fig9:c}}{\includegraphics[width=0.8in]{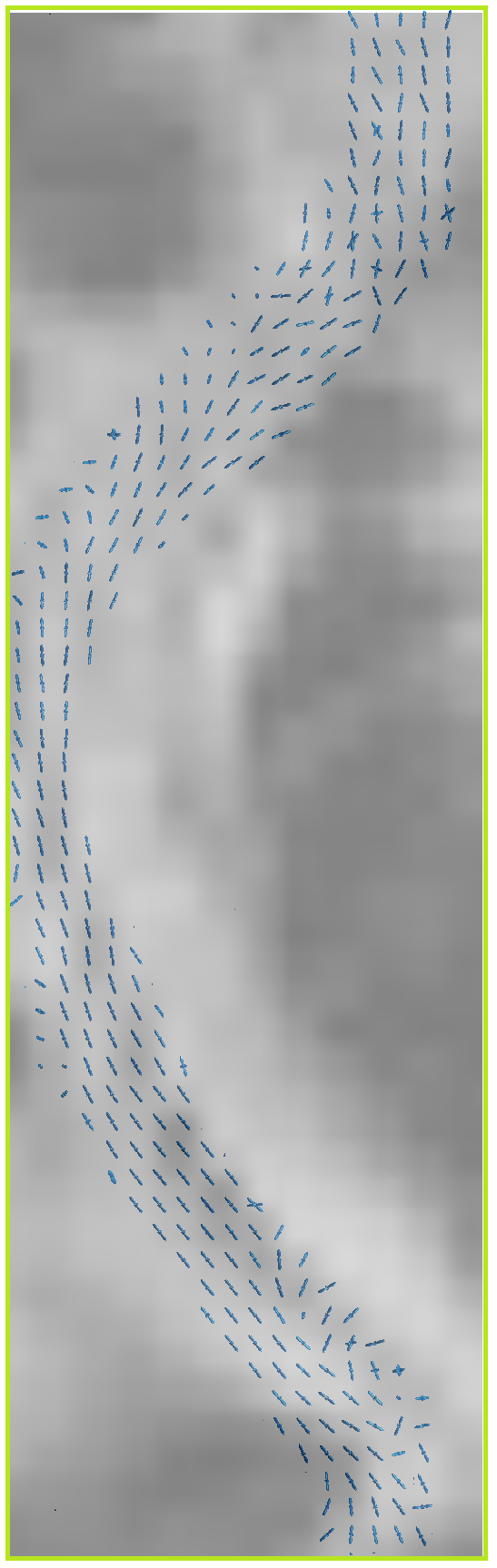}}
   
     \caption{(a) FA map of human brain MRI slice with blue box bordering the corpus callosum genu and yellow box bordering a part of corpus callosum splenium. (b)-(c) Fiber mapping image reconstructed using proposed algorithm for the
region of interest (highlighted using boxes) overlaid on the FA maps}
  \label{fig:9}
\end{figure}

Fig. \ref{fig9:a} shows the FA maps of the whole slice with regions
of corpus callosum genu and splenium enclosed in blue and green boxes. The reconstructed map corresponding to the corpus callosum genu and splenium region is shown in Fig. \ref{fig9:b} and Fog. \ref{fig9:c}, respectively. The bundles of straight, bending and crossing fibers are visible. The changing orientation and intersections of the fiber tracts are visible.

\section{Conclusion}
We developed a new methodology for reconstructing brain's axonal organization. This model is introduced to overcome the limitations of UGD and AGD models in estimating the orientation of fibers with higher accuracy. The proposed model is performing better in reducing angular error in both the cases: NNLS and OMP-TV2 slovers. Performance is more pronounced especially when the separation angle between the fibers is sufficiently small. In the real dataset, the proposed model successfully retraces the white matter fiber tracks. Performance of simulations at higher noise, $\sigma=0.1$ shows the stability of the proposed model at such level of noise. The real datasets are considered highly noisy due to multiple artefacts, hence proposed model is taken as a good choice for reconstruction purpose.

\section*{Declaration of Competing Interest}

The authors declare that they have no known competing financial interests or personal relationships that could have appeared to influence the work reported in this paper.

\section*{Acknowledgements}
One of the authors, Ashishi Puri, is grateful to the Ministry of Human Resource Development
India and the Indian Institute of Technology Roorkee for financial support through the The grant number MHR01-23-200-4028 to carry out this work. This work is also supported by the project grant no.
DST/INT/CZECH/P-10/2019 under Indo-Czech Bilateral Research Program.

\bibliographystyle{bst}
\bibliography{bib}
\end{document}